# Malware Squid: A Novel IoT Malware Traffic Analysis Framework Using Convolutional Neural Network and Binary Visualisation

Robert Shire[1], Stavros Shiaeles[1], Keltoum Bendiab[1], Bogdan Ghita[1], and Nicholas Kolokotronis[2]

[1] Centre for Security, Communications and Network Research,
University of Plymouth, Plymouth PL48AA, UK
Keltoum.bendiab@gmail.com, sshiaeles@ieee.or

[2] Department of Informatics and Telecommunications,
University of Peloponnese, 22131 Tripolis, Greece
nkolok@uop.gr

**Abstract**. Internet of Things devices have seen a rapid growth and popularity in recent years with many more ordinary devices gaining network capability and becoming part of the ever-growing IoT network. With this exponential growth and the limitation of resources, it is becoming increasingly harder to protect against security threats such as malware due to its evolving faster than the defence mechanisms can handle with. The traditional security systems are not able to detect unknown malware as they use signature-based methods. In this paper, we aim to address this issue by introducing a novel IoT malware traffic analysis approach using neural network and binary visualisation. The prime motivation of the proposed approach is to faster detect and classify new malware (zero-day malware). The experiment results show that our method can satisfy the accuracy requirement of practical application.

## I. Introduction

The explosive development of the concept of Internet of Things (IoT) is accompanied by an unprecedented revolution in the physical and cyber world. Smart, always- connected devices provide real-time contextual information with low overhead to optimize processes and improve how companies and individuals interact, work, and live. An increased number of businesses, homes and public areas are now starting to use these devices. The number of interconnected devices in use worldwide now exceeds 17 billion, number that is expected to grow to 10 billion by 2020 and 22 billion by 2025, according to a recent report [1].

On one side, the IoT devices offer extended features and functionality; on the other side, their security level is still low, with well-known weaknesses and vulnerabilities, such as easily guessable passwords and insecure default settings [2]. This gives cybercriminals the opportunity to easily exploit these vulnerabilities and create backdoors into a typical organisation infrastructure. To ensure their protection against potential vulnerabilities, these devices need to be updated and patched regularly; however, given they are not perceived as critical IT infrastructure, it is likely that they are less likely to be upgraded. Further, given their hardware, some IoT devices may not be patchable, and the only option is to replace them entirely when they become vulnerable [3]. Beyond the convenience or simplicity of patching, insecure Internet- connected IoT devices represent a security risk. According to a recent report of Symantec, IoT devices will increasingly represent an exploitation target; Symantec already found a 600% increase in overall IoT attacks in 2017 [4].

Botnets are the most common type of malware when an IoT device is compromised [5], either standalone or aggregated to become part of a botnet, capable of launching devastating DDOS (Distributed Denial of service) attacks. Given its uncommon architecture, once a botnet infects an IoT device, it can be very hard to detect the malware. Most conventional antimalware tools rely on a syntactic signature for their detection methods [6], where the signature of a file is compared to a list of known malicious ones. Thus, these systems all require a database with every known malware signature contained within it.

This is a very time-consuming process and requires already analysing the malware or its instruction sequence [7]. Moreover, the signature generation involves manual intervention and requires strict code analysis [6, 7], this pushes for enhanced, automated analysis. In this paper, we present a novel IoT malware traffic analysis method that addresses this issue by using a TensorFlow convolutional neural network paired with a binary visualization technique. The main contribution of this proposal is an automated malware traffic analysis method that combines binary visualisation of IoT traffic with the TensorFlow learning model. The combination is ideal for faster analysis of real-time traffic data compared to other approaches and makes it more appropriate to detect and analyse unknown zero-day malware. The proposal utilizes sockets to monitor devices network traffic, the Binvis binary data visualisation technique to convert the binary content of packets into 2D images, and the TensorFlow machine learning method to analyse the produced images. The objective of this analysis is to identify malware in the recent packets, based on the assumption that malware traffic tends to have a more clustered appearance of its patterns on the produced images whereas classic traffic presents more consistent and static. Obviously, both sides had anomalies and expectations.

The overall structure of the paper is organised as follows: Sect. 2 describes the prior works done in malware traffic analysis and classification. In Sect. 3, we present the methodology of the proposed method using neural network TensorFlow and binary visualization. Section 4 presents experiment results and analysis as well as a com- parison with other methods. Finally, Sect. 5 provides concluding remarks and future work.

## II. RELATED WORKS

Detection of malware and its associated traffic is still a persistent challenge for the security community. Research in this area is always needed to keep one step ahead of the hackers. However, IoT devices are upcoming new technology, especially inside a home environment, so anti-malware tools and associated research have been minimal compared to normal technologies [2]. Most attempts to detect or prevent malware traffic are performed by firewalls and intrusion prevention systems, for those in a home environment there is not much security other than the regular patches [8].

Several approaches have been proposed in the literature to detect or mitigate malware traffic. Signature-based detection techniques are the most used, however, they are unable to detect unknown malware traffic for which there exists no signature and involve manual interventions [6–8]. Machine learning is one of the most efficient techniques that have been employed to overcome this issue. Over the years, many machines learning approaches have been proposed for malware traffic analysis and classification. In [9], authors introduced the deep learning method of DBN (Deep Belief Networks) to the intrusion detection domain. In the proposed approach, authors used the DBN for malware traffic classification. Following the same direction, recent work in [10] proposed a malware traffic identification method using a sparse autoencoder. In this work, authors proposed a novel classifier model by combining the power of the Non-symmetric Deep Auto-Encoder (NDAE) (deep-learning), and the accuracy and speed of Random Forest (RF) (shallow learning), leading to high accuracy in malware detection. However, they both used a hand-designed flow features dataset as input data. On the other hand, Convolutional Neural Networks (CNN) and recurrent neural networks (RNN) are also used in many studies to perform malware traffic classification tasks based on spatial and temporal features. For example, authors in [11] transformed the network traffic features into a sequence of characters and then used RNNs to learn their temporal features. The RNN was then applied to detect malware traffic. While in this study the RNNs are used alone and learned a single type of traffic feature, authors in [12], authors used CNN to learn the spatial features of network traffic and achieved malware traffic classification using an image classification method. The proposed method needed no hand-designed features but directly took raw traffic as input data of the classifier, and the classifier then can learn features automatically. The CNN is then used to perform image classification of the images that were created from traffic sample PCAP files. This method has proven the efficiency of malware traffic classification using representation learning approach, being very successful in identifying classic traffic and even malware. However, it does not focus on unknown malware traffic. Thus, if the neural network is not already trained on the type of attack traffic, it will not be able to classify the traffic, or it will falsely categories it. Moreover, potentially missing Zero-day exploits traffic of viruses lets the work down as such threats are possibly the most serious ones to a network.

The study in [13] that also covered IoT intrusion detection used a different detection factor to help with the traffic analysis, more specifically the data associated with the CPU and memory usage of the IoT device. This is based on the observation that the CPU and memory usage tend to increase when a malware component is detected on the device. Although the CPU and memory features were effective, they require a lot of set up time and reconstruction of a testing network, making the method rather difficult to implement. In [14], the authors built a similar malware detection tool that focused on malware executables as opposed to traffic. This work also had analysis of binary visualisations through a neural network. The proposed approach uses binary visualisation to convert a binary file data into an image, and self-organizing incremental neural networks (SOINN) for the analysis and detection of malicious payloads. The limitations of this work stemmed from the limited availability of samples, leading to restricting neural network training options.

## III. THE PROPOSED METHOD

The proposed IoT malware traffic analysis method consists of three main steps, as shown in Fig. 1. The first step is the network traffic collection, through either directly sniffing the network or using files containing pre-captured network traffic that can be replayed through TCPreplay for the sniffer to collect again. The second step is the binary visualisation phase, which takes the collected traffic stored in ASCII (American Standard Code for Information Interchange) and convert it into a 2D image. In the final step, the binary image is then processed by the TensorFlow module, which analyses it against its training modules.

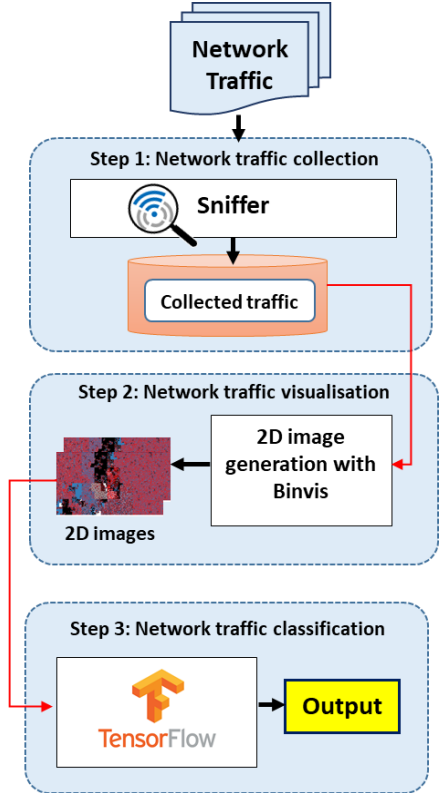

**Fig. 1**. Overview of the proposed method.

### A. NETWORK TRAFFIC COLLECTION

Packet capturing is the most used scheme to accomplish the goal of network data collection [15]. Typically, packet-based collection mechanisms use sniffers to implement network data collection through centralized management such as Wireshark, nmap, Airodump, and TCPdump. A sniffer is regarded as a convenient and efficient tool to detect traffic and capture packets [15]. In our approach, we proposed a network traffic collection method using a Python-based tool [16] which ensures two major tasks are accomplished: collection and storage.

1) **Traffic collection**: When the sniffer is loaded into memory, it can collect all packets that are either traversing the network or are replayed. The used Sniffer utilizes Python sockets at a low-level networking interface to collect packets. It is worth noting that the proposed approach is also applicable for the case of very sporadic IoT traffic as it creates profile of what is normal traffic and compare it with abnormal.

2) **Traffic storage**: Received data is passed out to a file that contains the data from the payload in the packet, this data is turned to hexadecimal so that Binvis can plot it into a 2D image in the second step.

The dataset is created by using two main collection methods, one that the sniffer collected of both normal traffic and malware traffic and that used traffic from pcap files. All the collected files came from real-world network environments rather than be artificially generated data. If traffic samples have a large size, only parts of the PCAP file is used. Similarly, traffic samples that are too small will only be used on a slower TCPreplay speed. The collected pcap-based files are added to the dataset and replayed through the module TCPreplay using various speeds.

### B. TRAFFIC VISUALISATION

In this work, we use a visual representation algorithm of the traffic collected that is based on Binvis [17]. This binary data visualization tool converts the contents of a binary file to another domain that can be visually represented (typically a two- dimensional space) [14]. Binvis represents the different ASCII values by using red, green and blue colour classes as shown in Table 1, while black (0x00) and white (0xFF) classes are used to represent null and (non-breaking) spaces.

Table 1. Binvis colour divisions

| Colour | Division |
| --- | --- |
| Blue | if the ASCII character is printable |
| Green | if the character is control |
| Red | if the character is extended ASCII |
| Black | 0x00 |
| White | 0xFF |

To convert a binary file into a 2D image, its data are seen as a byte string, where each byte value is compared against the ASCII table and is attributed to a colour according to the division it belongs, as outlined in Table 1. In our approach, the binary file is made from network packets collected by the sniffer, these are then converted into a string of hexadecimal characters which is later used to create the image of the traffic, using a clustering algorithm. The final output of Binvis is an image that represents the features of network traffic. The Hilbert space-filling curve clustering algorithm is used in Figs. 3 and 4. This algorithm surmounts other curves in preserving the locality between objects in multi-dimensional spaces [14, 18], thus creating a much more appropriate imprint of the image. This helps the machine learning neural network analysis the image for anomalies in normal traffic.

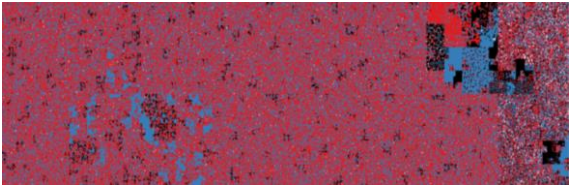

**Fig. 3**. Binary visualisation of Botnet pcap file

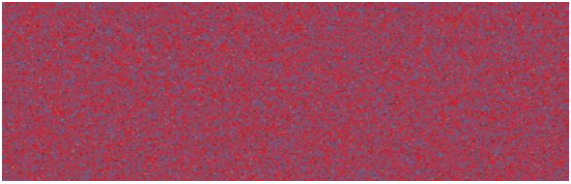

**Fig. 4**. Binary visualisation of normal pcap file

### C. MALWARE TRAFFIC ANALYSIS

In this step, TensorFlow is used to analyse the produced images against its in-depth training. TensorFlow is a machine learning system that operates at large scale and in heterogeneous environments [19], providing full flexibility for implementing any type of model architecture [20]. Moreover, TensorFlow is an effective machine learning algorithm to analyse images and classify them; accordingly, it is easy to retrain and learns quickly from updates to the neural network [19]. Its outstanding artificial intelligence feature is its excellent image recognition ability, which is specifically why it is being utilised within this application. The TensorFlow AI could easily detect differences between the images, including differences that the human eye could not detect [20].

The TensorFlow module utilizes a CNN which works like a classic neural network but has an extra layer at the beginning called the convolution. The binary output from Binvis is broken up into a number of tiles and, while the machine learning aims to predict what each tile is, the AI then aims to determine the combination of tiles that the picture is based on. This allows TensorFlow to parallelize operations and detect the object regardless of where it is located in the image [21].

The machine learning process is separated into two stages. The first stage is the training phase, where the MobileNet module is employed for the retraining element [21]. MobileNet is a neural network that is very small and efficient, chosen for its lightweight element. It is designed specifically to be mindful of the resources it takes up on a device or application [21]. In the second stage, the image files are tested against the samples of the database to perform classification.

### IV. EXPERIMENTAL RESULT

In this section, we present the performance analysis results of the prototype that was implemented based on the methodology presented in the previous section.

Accuracy (A), precision (P), recall (R) and f1 value (F1) metrics were used to evaluate the overall performance of the proposed malware traffic classification approach.

$$A = \frac{TP + TN}{TP + FP + TN + FN} \quad (1)$$

$$P = \frac{TP}{TP + FP} \quad (2)$$

$$R = \frac{TP}{TP + FN} \quad (3)$$

$$F1 = \frac{2 \times P \times R}{P + R} \quad (4)$$

Where TP is the number of instances correctly classified as good traffic, TN is the number of instances correctly classified as bad traffic, FP is the number of instances incorrectly classified as good traffic, and FN is the number of instances incorrectly classified as bad traffic.

### A. EXPERIMENT SETUP

The simulation experiments were performed on a virtual machine built on VM workstation, running Ubuntu 18.0.4. The ISO was not updated during this time to keep prevent technology incompatibility. The dataset that was used in testing, contained a set of 100 pcap samples, collected from external repositories. It is composed of a mixture of 30 normal and 70 malware traffic samples. Samples were classified into the unknown section if the pcap collected was unnamed or Wireshark testing came back inclusive. Table 2 summarizes the percentage of malicious traffic samples of the whole data set.

Table 2. Malicious traffic sample percentage according to type of malware

| Malware type | Percentage |
| --- | --- |
| Trojans | 25% |
| DDoS | 16% |
| Botnets | 19% |
| OS scan | 8% |
| Keylogger | 6% |
| Backdoors | 10% |

In the training stage, the TensorFlow algorithm was trained by 500 iterations of the data set in a static environment. Knowing that the more training, the better the accuracy of traffic analysis and classification, however incorrect training samples will result in a flawed neural network that can only produce inaccurate results. The minimum training requirements are 30 images for each section, which was 30 images for normal traffic and 30 for malware traffic. Since the TensorFlow was unable to detect whether traffic was good or bad at this stage, the samples used in this stage had to be labelled as being good or malware traffic. For the testing process, the set up was a home scenario, with the thermometer was chosen for the malware host because it is one of the most common home IoT devices and in recent years has been responsible for some of the most devastating attacks [23]. Therefore, it was fitting to use it as the testing scenario. Collected sample files were replayed using TCPreplay on the same network interface card to homogenize the network behavior exhibited by the datasets.

### B. EXPERIMENTS RESULTS AND ANALYSIS

Several tests were carried out to determine the accuracy of the proposed classifier after the addition of more samples, in each test more samples were added to the training data for the machine learning to be retrained on. Figure 5 shows the average accuracy of traffic in the different tests. Four set tests were completed, where the fourth test is the final test with the most training samples that were collected being used. It is apparent from the four tests that the overall accuracy rate of normal traffic stayed consistent throughout, only varying from 78% to 90% because most normal traffic was found to have very similar characteristics throughout the data stream. This makes the classification of this traffic very easy since the training data would almost always be able to match the sample pcap files against the current traffic being tested. However, even from the start of testing the good traffic had a high accuracy rate, which was surprising given the number of samples in the training at that point. This is the result of test 1 which has a high false positive rate. Knowing that, during early testing, if the algorithm did not recognise a binary visualisation, then it would classify it as good traffic, leading to the 60% - 40% split for malware traffic (*see* Fig. 5).

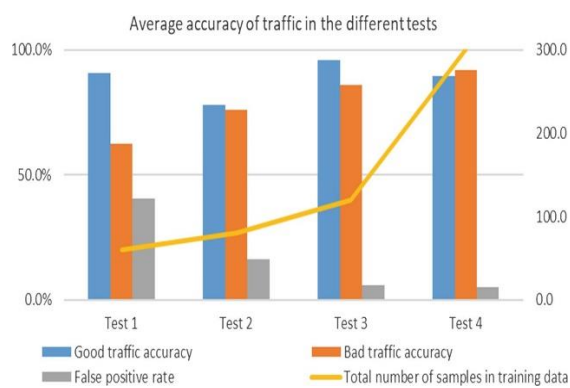

**Fig. 5**. Average accuracy of malware and good traffic throughout the 4 sets of tests with comparison to number of training data used.

As shown in Fig. 6, the issue with the high false positive rate was slowly being phased out, with the addition of more samples to the neural network throughout the tests, dropping from 40% to 5% by the final set of tests. The addition of more training data mainly contained more specific types of malware samples, in the first and second tests, one pcap of a Trojan was used to train the neural network, while no backdoor attacks were used. This made it near impossible for the algorithm to detect more of these types of traffic without more in-depth training. Malware traffic accuracy varied massively across all the tests, starting at a low accuracy (60%) but by the final test ended up reaching a good accuracy of 91%. Figure 6 shows the final stage of testing results.

The stage has been broken down to show the accuracy of the individual types of traffic, DDOS traffic had clearly the best accuracy rate by the end test which even though it made up only 16% of the overall traffic it had a clear pattern where its malware samples were mainly covered in green pixels. This indicates the use of the control character being used over the staple amount, which the machine learning has clearly learned this malicious trait over four sets of tests and uses it to classify DDOS attacks. The algorithm had a much higher probability of showing false positives than false negatives, thus false negative data was not added to the graph results. In summary, the amount of training was the variable that had the most effect on the accuracy of the neural networks.

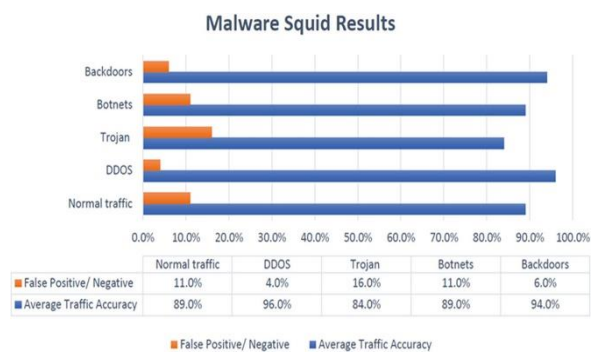

**Fig. 6**. Final test individual average accuracy for each malware type.

Table 3 shows that the proposed approach achieved an accuracy of 91.32%, which meets accuracy of practical use. It has got a high precision 91.67% and recall (91.03%), which shows the ability of our approach in classifying bad and good traffic.

Table 3. Results for the last test

|        | (A)    | (P)    | (R)    | (F1)   |
|--------|--------|--------|--------|--------|
| Test 4 | 91.32% | 91.67% | 91.03% | 91.35% |

### C. ASCII Characters Frequency Throughout Traffic Results and Analysis

As can be seen in Fig. 7, the different types of malwares have distinctive features to differentiate them. Whereas normal traffic can be spotted by their more even distribution of ASCII characters or colours across an image, most of the malware samples follow the same pattern of having more predominance of black (Null Bytes) or white areas (Spaces) in their samples, however, the DDOS is an exception with its extremely high frequency of Control characters. Malware samples do not follow the same pattern as normal traffic, the large volume of null and white spaces might indicate that code was present in the traffic stream. Null bytes which are normally used in coding to mark the end of the string or its termination point [24], could indicate the use of traffic containing a back-door attack or similar.

Null bytes are also the main factor in injection exploitation techniques used to bypass security filters, the null bytes are added to user- supplied data to manipulate application behaviour that called a null byte injection attack [25]. Null bytes are also commonly not contained within the default ASCII web request [26], an indication of potential botnet usage that is targeting web servers with no intention to establish a legit connection. DDOS attacks also had an interesting pattern that did not match up with the rest of the other malware, as displayed in Fig. 7, the images had a very high frequency of green pixels, more than any type of traffic recorded. High levels of green pixels represent an abuse of the use in control characters [14], attacks commonly use control characters to hide data in packets that are malicious in nature [25].

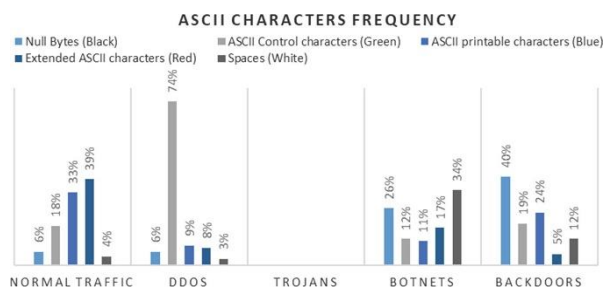

**Fig. 7**. Average ASCII character frequency between malware and normal traffic PCAPS.

### D. Comparison

It is not easy to conduct a fair comparison among various malware classification approaches due to the differences between the datasets of traffic used, image visitation tools used and target environments. Thus, our comparison will be based on some significant features. Table 4 overviews a general comparison between our approach and the well-known IDS (Intrusion Detection System) Snort [28] and Suricata [29].

Table 4. Comparison with other methods

| Features             | Malware squid | Snort | Suricata |
|----------------------|---------------|-------|----------|
| Low false alarm rate | Yes           | No    | Yes      |
| Lightweight          | Yes           | Yes   | No       |
| Protocol independent | Yes           | No    | No       |
| Raw traffic input    | Yes           | Yes   | Yes      |

The results from the experiments show that malware squid has a low false alarm rate, only beaten by Suricata due to its more modern nature and detection methods. Whereas snort has a high false alarm rate due to problems with extracting malware footprints from traffic, the means of which its Snort rule set runs off [12]. Malware squid is also a lightweight program for one that utilises an AI, this is due to the MobileNet algorithm being as minimalistic as possible, which is also similar to the older Snort [29], however Suricata Is not lightweight due to its increased memory consumption used in multithreading [30]. Both methods use a set rule set to detect malicious traffic, if traffic matches these sets it will trigger an alarm [12], Malware squid uses image classification so has no knowledge of rule sets making it protocol independent. Finally, all three approaches can take raw traffic input into their datasets [27], this seems to be a staple in IDS detection technologies.

### V. Conclusion

This paper proposed a novel IoT malware traffic analysis method, leveraging multi- level artificial intelligence that uses a combination of neural network paired with a binary visualization. The method can be used to protect IoT devices on gateway level bypassing the limitations associated with the IoT environment. From our initial experimental results, the method seems promising and being able to detect unknown malware. Moreover, the method learns from the misclassifications and improve its efficiency. Future work would involve the use of more samples for training and testing and utilising GPU for binary visualization and CNN classification and testing the proposed approach for encrypted traffic as well.

**ACKNOWLEDGEMENT.**

This project has received funding from the European Union's Horizon 2020 research and innovation programme under grant agreement no. 786698. This work reflects authors' view and Agency is not responsible for any use that may be made of the information it contains.